\documentclass{revtex4}
\usepackage{epsfig}
\usepackage{amssymb}
\usepackage{amsmath}
\usepackage{amsfonts}
\usepackage{graphicx}
\usepackage{mathrsfs}
\usepackage[dvips]{color}
\usepackage{multirow}

\newcommand{\R}{\mathbb{R}}

\newcommand{\fn}{{\,\mathfrak{n}\,}}

\newcommand{\fz}{\mathfrak{z}}

\newcommand{\cB}{\mathcal{B}}

\newcommand{\cD}{\mathcal{D}}
\newcommand{\cH}{\mathcal{H}}
\newcommand{\cE}{\mathcal{E}}

\newcommand{\cO}{\mathcal{O}}

\newcommand{\be}{\begin{equation}}
\newcommand{\ee}{\end{equation}}
\newcommand{\bea}{\begin{eqnarray}}
\newcommand{\eea}{\end{eqnarray}}
\newcommand{\nn}{\nonumber}
\newcommand{\kt}{\rangle}
\newcommand{\br}{\langle}

\newcommand{\ed}{\end{document}}

\newcommand{\bi}{\begin{itemize}}
\newcommand{\ei}{\end{itemize}}

\newcommand{\bce}{\begin{center}}
\newcommand{\ece}{\end{center}}

\newcommand{\sE}{\mathscr{E}}

\newcommand{\sH}{\mathscr{H}}

\newcommand{\sT}{\mathscr{T}}
\newcommand{\RE}{{\rm Re}}
\newcommand{\IM}{{\rm Im}}

\begin{document}

\title{Spectral Singularities in the Surface Modes of a Spherical Gain Medium}

\author{Ali~Mostafazadeh and Mustafa~Sar{\i}saman}
\address{Department of Mathematics, Ko\c{c}
University,\\ Sar{\i}yer 34450, Istanbul, Turkey\\
amostafazadeh@ku.edu.tr}

\begin{abstract}
We study the surface modes of a homogenous spherical gain medium and provide a comprehensive analytic treatment of a special class of these modes that support spectral singularities. Because the latter have a divergent quality factor, we call them the singular gallery modes. We show that they can be excited using arbitrarily small amounts of gain, and as a result, the system lacks a lasing threshold, effectively. This shows that we can realize spectral singularities in the surface modes of extremely small spherical samples with modest amounts of gain. We also examine the possibility of exciting singular gallery modes with different wavelengths using the same amount of gain. This corresponds to the situation where the system undergoes simultaneous lasing at different wavelengths.
\medskip

\noindent {Pacs numbers: 42.25.Bs, 42.60.Da, 03.65.Nk, 24.30.Gd}
\end{abstract}

\maketitle

\section{Introduction}

Physical meaning and realizations of spectral singularities \cite{ss} have attracted much attention in recent years \cite{prl-2009} -- \cite{pra-2013b}. In particular, it is shown that in the realm of optics they correspond to the lasing at the threshold gain \cite{pra-2011a} and that a time-reversed optical spectral singularity is responsible for the coherent perfect absorption of light \cite{longhi-phys-10}.

In Ref.~\cite{pla-2011} we study the optical spectral singularities (OSSs) of a uniform spherical gain medium and show that the emergence of an OSS puts a lower bound on the radius of the gain medium. In Ref.~\cite{prsa-2012} we show that this lower bound can be reduced by coating the medium with a high-refractive-index material. It turns out, however, that for typical gain media such as the dye laser material examined in Refs.~\cite{pla-2011,prsa-2012} such a coating is not sufficient to reduce the critical size for supporting spectral singularities to the micrometer range. This in turn leads to obvious problems related with maintaining a substantial amount of uniform gain throughout the sample.

The main reason for our inability to create OSS for smaller samples is that in Refs.~\cite{pla-2011,prsa-2012} we consider OSS that arise in the radial modes of the sphere. The idea that this problem can be avoided by considering OSS in the surface modes provides the main motivation for the present study. As a first step in this direction we have explored in Ref.~\cite{pra-2013b} the OSS in the surface modes of an infinite cylindrical gain material. These modes that we call ``singular gallery modes'' share most of the characteristic features of the well-known whispering gallery modes, but unlike the latter have a divergent quality factor. In the present article we extend the analysis and results of \cite{pra-2013b} to a homogeneous spherical gain medium which is practically more important and mathematically more involved.

\section{Multipole Solutions of Maxwell's Equations}

Consider an optically active material of spherical shape. Let $a$ and $\fn$ respectively denote its radius and complex refractive index, and suppose that the latter is independent of space and time. The Maxwell equations describing the interaction of the electromagnetic waves with this system have the form:
    \begin{align}
    &\vec{\nabla}\cdot\vec{\cD} = 0, &&
    \vec{\nabla}\cdot\vec{\cB} = 0,
    \label{equation1}\\
    &\vec{\nabla} \times \vec{\cE} + \dot{\vec{\cB}} = 0, &&
    \ \vec{\nabla} \times \vec{\cH} - \dot{\vec{\cD}} = 0,
    \label{equation2}
    \end{align}
where $\vec{\cD}(\vec r):= \varepsilon_0\fz(r) \vec{\cE}(\vec r)$,
$\vec{\cH}(\vec r):= \mu_0^{-1} \vec{\cB}(\vec r)$,  $\varepsilon_0$ and $\mu_0$ are respectively the permeability and permittivity of the vacuum, $\vec r$ is the position vector, $r:=|\vec r|$ is the radial spherical coordinate,
    \be
    \fz(r):=\left\{\begin{array}{ccc}
    \fn^2 & {\rm for} & r<a,\\
    1 & {\rm for} & r\geq a,
    \end{array}\right.
    \label{e1}
    \ee
and an over-dot denotes a time-derivative.

Assuming a harmonic time-dependence for the fields, $\vec{\cE}(\vec r)=e^{-i\omega t}\vec{E}(\vec r)$ and $\vec{\cH}(\vec r)=e^{-i\omega t}\vec{H}(\vec r)$, we can use (\ref{equation2}) to obtain
    \begin{align}
    &\left(\nabla^{2} +\tilde{k}^{2}\right) \vec{E}(\vec{r}) = 0, &&
    \left(\nabla^{2} +\tilde{k}^{2}\right) \vec{H}(\vec{r}) = 0,
    \label{equation4}\\
    &\vec{H}(\vec{r}) = -\frac{i}{k Z_{0}} \vec{\nabla} \times \vec{E}(\vec{r}), &&
    \vec{E}(\vec{r}) = \frac{i Z_{0}}{k \fz(r)} \vec{\nabla} \times \vec{H}(\vec{r}),
    \label{equation5}
    \end{align}
where $k := \omega / c$ is the wavenumber, $c := 1 / \sqrt{\mu_{0}
\epsilon_{0}}$ is the speed of light in vacuum, $Z_{0} :=
\sqrt{\mu_{0}/\epsilon_{0}}$ is the impedance of the vacuum, and
$\tilde{k} := k \sqrt{\fz(r)}$.

Following the standard treatment of the vector Helmholtz equations (\ref{equation4}), we construct the multipole solutions of these equations using the vector spherical harmonics:
      \be
      \vec{X}_{\ell m}:=\frac{-i\,\vec{r} \times \vec{\nabla}Y_{\ell m}(\theta, \varphi)}{\sqrt{\ell (\ell + 1)}},
      \label{equation17}
      \ee
where $\ell$ is a non-negative integer, $m$ is an integer fulfilling $|m|\leq\ell$, and $Y_{\ell m}$ are the standard spherical harmonics \cite{jackson}. This yields the transverse electric (TE) and transverse magnetic (TM) multipole solutions of (\ref{equation4}) that are respectively given by
    \begin{align}
    &\vec{E}_{\ell m}^{(E)}=\sE(r)  \vec{X}_{\ell m},
    && \vec{H}_{\ell m}^{(E)} = -\frac{i}{Z_{0}k} \vec{\nabla} \times \left[\sE(r)  \vec{X}_{\ell m}\right],
    \label{equation18}\\
    &\vec{E}_{\ell m}^{(M)} = \frac{iZ_{0}}{k \fz(r)} \vec{\nabla} \times \left[\sH(r) \vec{X}_{\ell m}\right],
    &&\vec{H}_{\ell m}^{(M)} = \sH (r) \vec{X}_{\ell m}.
    \label{equation19}
    \end{align}
Here $\sE$ is a scalar function of the form
    \be
    \sE(r):=\left\{\begin{array}{ccc}
    a_0\,j_\ell (k\fn r) & {\rm for} & r<a,\\
    a_1\, h_\ell^{(1)}(k r) + a_2\, h_\ell^{(2)}(k r) & {\rm for} & r\geq a,
    \end{array}\right.
    \label{equation9}
    \ee
$a_i$, with $i=0,1,2$, are complex coefficients possibly depending on $\ell$ and $m$, $j_\ell$ and $h_\ell^{(1,2)}$ are respectively the spherical Bessel and Hankel functions, and $\sH$ is a scalar function having the same form as $\sE$ (with a possibly different choice for $a_i$.)

Notice that the solutions (\ref{equation18}) and (\ref{equation19}) are acceptable provided that they satisfy the appropriate boundary conditions for the problem, namely that at $r=a$ the tangential component of both the $\vec E$ and $\vec H$ fields must be continuous functions of $r$. Table~\ref{table01} gives explicit expressions for the components of the electric and magnetic fields in the spherical coordinates, and Table~\ref{table02} shows the equations representing the boundary conditions satisfied by the electric and magnetic fields. Here for brevity, we do not make the $\ell$- and $m$-dependence of the field components explicit and absorb a common constant factor in the definition of the coefficients $a_i$ appearing in the expression for $\sE$ and $\sH$. We also introduce $x:=k a$ and
    \be
    T^{(1)}_{\ell m}(\theta):=\frac{m P^m_{\ell}(\cos\theta)}{\sin\theta},~~~~~
    T^{(2)}_{\ell m}(\theta):=\partial_\theta P_{\ell}^{m} (\cos\theta),
    \label{T1-2}
    \ee
where $P_{\ell}^{m}$ are the associated Legendre polynomials, and for every differentiable function $f(r)$,
    \be
    \tilde f(r):=\left(\frac{d}{dr}+\frac{1}{r}\right)f(r).
    \label{tilde=}
    \ee

    \begin{table}
    \begin{center}
	{
    \begin{tabular}{|c|c|c|c|c|c|}
    \hline
    TE-Fields & TM-Fields \\
    \hline & \\[-5pt]
    $\begin{aligned}
    & E_{r}= 0\\[3pt]
    & E_{\theta}=- \sE(r)\,T^{(1)}_{\ell m}(\theta)\,e^{im\varphi}\\[3pt]
    & E_{\varphi} =- i \sE(r)\,T^{(2)}_{\ell m}(\theta)\,e^{i m \varphi}\\[3pt]
    &H_{r} =\frac{\sE(r)}{Z_0kr}\,P^m_{\ell}(\theta)\,e^{i m \varphi}\\[3pt]
    &H_{\theta} =\frac{\tilde{\sE}(r)}{Z_0k}\,T^{(2)}_{\ell m}(\theta)\,e^{i m \varphi}\\[3pt]
    & H_{\varphi} = \frac{i \tilde{\sE}(r)}{Z_0k}\,T^{(1)}_{\ell m}(\theta)\,e^{i m \varphi}\\[3pt]
    \end{aligned}$ &
    $\begin{aligned}
    & E_{r}=-\frac{Z_{0}\sH(r)}{rk\fz(r)}\,P^m_{\ell}(\theta)\,e^{i m \varphi}\\[3pt]
    & E_{\theta} =- \frac{Z_{0}\,\tilde{\sH}(r)}{k\fz(r)}\,T^{(2)}_{\ell m}(\theta)\,e^{i m \varphi}\\[3pt]
    & E_{\varphi} =- \frac{i Z_{0}\,\tilde{\sH}(r)}{k \fz(r)}\,T^{(1)}_{\ell m}(\theta)\,e^{i m \varphi}\\
    &H_{r} =0\\[3pt]
    &H_{\theta} = -\sH(r)\,T^{(1)}_{\ell m}(\theta)\,e^{i m \varphi}\\[3pt]
    &H_{\varphi} =-i \sH(r)\,T^{(2)}_{\ell m}(\theta)\,e^{i m \varphi}\\[-8pt]
    &
    \end{aligned}$\\
    \hline
    \end{tabular}}
    \vspace{6pt}
    \caption{Components of the TE and TM fields in spherical coordinates.}
    \label{table01}
    \end{center}
    \end{table}

    \begin{table}[t]
    \begin{center}
	{
    \begin{tabular}{|c|c|c|c|c|c|}
    \hline
    TE-Fields & TM-Fields \\
    \hline
    & \\[-8pt]
    $\begin{aligned}
    & a_0 j_{\ell}(\fn x) = a_1 h_{\ell}^{(1)} (x) + a_2 h_{\ell}^{(2)} (x)\\[3pt]
    & a_0 \fn \tilde{j}_{\ell}(\fn x) = a_1 \tilde{h}_{\ell}^{(1)} (x) + a_2 \tilde{h}_{\ell}^{(2)}(x)\\[3pt]
    \end{aligned}$
    &
    $\begin{aligned}
    & a_0 j_{\ell}(\fn x) = a_1 h_{\ell}^{(1)} (x) + a_2 h_{\ell}^{(2)} (x)\\[3pt]
    & \frac{a_0}{\fn} \tilde{j}_{\ell}(\fn x) = a_1 \tilde{h}_{\ell}^{(1)} (x) + a_2 \tilde{h}_{\ell}^{(2)} (x)\\[3pt]
    \end{aligned}$\\
    \hline
    \end{tabular}}
    \vspace{6pt}
    \caption{Boundary conditions for TE and TM fields. Here $x:=k a$.}
    \label{table02}
    \end{center}
    \end{table}

According to Table~\ref{table02} the reflection amplitude for the TE and TM waves are given by \cite{pla-2011}
    \be
    R:=\frac{a_1}{a_2}=\left\{\begin{array}{cc}
    \displaystyle\frac{\fn h_\ell^{(2)}(x)\, \tilde{j}_{\ell}(\fn x)-\tilde{h}_{\ell}^{(2)}(x)\, j_\ell (\fn x)}{\tilde{h}_{\ell}^{(1)}(x)\, j_\ell(\fn x) - \fn h_\ell^{(1)}(x)\, \tilde{j}_{\ell}(\fn x)}
    & \mbox{for TE waves},\\[18pt]
    \displaystyle\frac{h_\ell^{(2)}(x)\, \tilde{j}_{\ell}(\fn x)-\fn\tilde{h}_{\ell}^{(2)}(x)\, j_\ell(\fn x)}{\fn\tilde{h}_{\ell}^{(1)}(x)\,j_\ell(\fn x)- h_\ell^{(1)}(x)\tilde{j}_{\ell}(\fn x)} & \mbox{for TM waves}.\end{array}\right.
    \label{TM-reflection}
    \ee
Notice that in view of (\ref{tilde=}) the above expression for the reflection amplitude of the TE waves simplifies, and we find
    \be
    R=\frac{\fn h_\ell^{(2)}(x)\,{j}'_{\ell}(\fn x)-{h}_{\ell}^{(2)\prime}(x)\,j_\ell(\fn x)}{
    {h}_{\ell}^{(1)\prime}(x)\,j_\ell(\fn x)-\fn h_\ell^{(1)} (x)\,{j}'_{\ell}(\fn x)}
    ~~~~\mbox{for TE waves}.
    \label{TE-reflection}
    \ee
Here a prime denotes the derivative of the corresponding function.

Spectral singularities are given by the real values of $k$ at which the reflection amplitude diverges. According to (\ref{TM-reflection}) and (\ref{TE-reflection}), they correspond to the real values of $k$ (and hence $x$) that satisfy
    \bea
    {h}_{\ell}^{(1)\prime}(x)\,j_\ell(\fn x)-\fn h_\ell^{(1)} (x)\,{j}'_{\ell}(\fn x)&=&0~~~~\mbox{for TE waves},
    \label{SSE-eq}\\
    \fn\tilde{h}_{\ell}^{(1)}(x)\,j_\ell(\fn x)- h_\ell^{(1)}(x)\tilde{j}_{\ell}(\fn x)&=&0~~~~\mbox{for TM waves}.
    \label{SSM-eq}
    \eea

\section{Whispering and Singular Gallery Modes}

Whispering gallery models (WGMs) are field configurations that propagate in a close vicinity of the inner boundary of a cylindrical or spherical medium \cite{rayleigh}. For our spherical model, Eqs.~(\ref{equation18}) and (\ref{equation19}) give an infinite family of exact solutions of the wave equation. In order to see if this family includes WGMs, we examine the corresponding time-averaged energy density and Poynting vector which are respectively given by
     \begin{align}
     &\br u \kt:=\frac{1}{4}\,\RE\left(\vec E\cdot\vec D^* + \vec B\cdot\vec H^*\right),
     &&\br\vec S\kt:=\frac{1}{2}\,\RE\left(\vec E\times\vec H^*\right).
     \label{u-S}
     \end{align}
In the following we study the behavior of these quantities for the multipole TE and TM field configurations, separately.

\subsection{TE Waves}

Employing Eq.~(\ref{u-S}) and the formulas listed in Table~\ref{table01} for the components of the TE fields, we find
    \begin{align}
	&\br u\kt=\frac{\varepsilon_0|\sE(r)|^2 \sT_{0}(\theta)^2}{4}\left\{\RE[\fz(r)]+\frac{1}{k^2r^2}\left[
    \left|\frac{r{\sE}'(r)}{\sE(r)}+1\right|^2+\sT_{1}(\theta)^2\right]\right\},
    \label{TE-u=}\\
    &\br\vec S\kt=\frac{|\sE(r)|^2 \sT_{0}(\theta)^2}{2Z_0kr}\left\{
    \IM\left[\frac{{r\sE}'(r)}{\sE(r)}\right]\hat r+\sT_{2}(\theta)\,\hat\varphi\right\},
    \label{TE-S=}
    \end{align}
where
	\bea
    \sT_{0}(\theta)&:=&\sqrt{T^{(1)}_{\ell m}(\theta)^2+T^{(2)}_{\ell m}(\theta)^2}=
    \sqrt{\big[\partial_\theta P^m_\ell(\cos\theta)\big]^2+\frac{m^2P^m_\ell(\cos\theta)^2}{\sin^2\theta}},
    \label{T0=}\\
    \sT_{1}(\theta)&:=&\frac{P^m_{\ell}(\cos\theta)}{\sT_{0}(\theta)}
    =\frac{\sin(\theta)
    P^m_\ell(\cos\theta)}{\sqrt{\sin^2(\theta)\left[\partial_\theta P^m_\ell(\cos\theta)\right]^2+m^2P^m_\ell(\cos\theta)^2}},
    \label{T1=}\\
    \sT_{2}(\theta)&:=&\frac{T^{(1)}_{\ell m}(\theta)P^m_{\ell}(\cos\theta)}{\sT_{0}(\theta)^2}=\frac{m \sin(\theta)P^m_\ell(\cos\theta)^2}{\sin^2(\theta)\left[\partial_\theta P^m_\ell(\cos\theta)\right]^2+m^2P^m_\ell(\cos\theta)^2}.
    \label{T2=}
    \eea

Inside the sphere, where $r<a$, $\fz(r)=\fn^2$, $\sE(r)=a_0 j_\ell(k\fn r)$, and Eqs.~(\ref{TE-u=}) and (\ref{TE-S=}) take the form
    \begin{align}
	&\br u\kt=\frac{\varepsilon_0\big|a_0 j_\ell(k\fn r)\big|^2 \sT_{0}(\theta)^2}{4}
	\left\{\RE(\fn^2)+\frac{1}{k^2r^2}\left[
    \left|\frac{k\fn r\ {j}'_\ell(k\fn r)}{j_\ell(k\fn r)}+1\right|^2+\sT_{1}(\theta)^2\right]\right\},
    \label{TE-u=in}\\
    &\br\vec S\kt=\frac{\big|a_0 j_\ell(k\fn r)\big|^2 \sT_{0}(\theta)^2}{2Z_0 k r}
    \left\{k r \,\IM\left[\frac{\fn \:{j}'_\ell(k\fn r)}{j_\ell(k\fn r)}\right]\hat r+\sT_{2}(\theta)\,\hat\varphi\right\}.
    \label{TE-S=in}
    \end{align}
As a result, the angle between $\br\vec S\kt$ and the tangent plane to the surface of the sphere is given by
    \be
    \Theta:=\tan^{-1}\left[\frac{\br\vec S\kt \cdot (-\hat r)}{\br\vec S\kt\cdot\hat\theta+\br\vec S\kt\cdot\hat\varphi}\right]
    =-\tan^{-1}\left\{\frac{k r}{\sT_{2}(\theta)}\IM\left[\frac{\fn {j}'_\ell(k\fn r)}{j_\ell(k\fn r)}\right]\right\},
    \label{Theta=01}
    \ee
where $\vec v\cdot\vec w$ stands for the standard dot product of vectors $\vec v$ and $\vec w$ in the Euclidean space $\R^3$.

Equations~(\ref{TE-u=in}) -- (\ref{Theta=01}) have the same structure as the corresponding equations  for the TE waves associated with the cylindrical gain medium of Ref.~\cite{pra-2013b}. In order to gain a better understanding of the consequences of (\ref{TE-u=in}) and (\ref{Theta=01}), we express them in terms of the real and imaginary parts, $\eta$ and $\kappa$, of $\fn$ and, in view of the fact that $|\kappa|\ll 1<\eta$, expand the result in powers of $\kappa$. In this way we find
     \begin{align}
     \br u\kt &= \frac{\varepsilon_{0}\left|a_0\right|^2 \sT_{0}(\theta)^2 \eta^2 F_+(\zeta)}{4} +\mathcal{O}(\kappa^2),
	\label{avrg-u=02}\\
     \Theta &= \tan^{-1}\left\{\frac{\kappa}{\eta \sT_{2}(\theta)}\frac{\zeta^2 F_{-}(\zeta)}{j^2_{\ell}(\zeta)}\right\}+\mathcal{O}(\kappa^2),
     \label{Theta=02}
     \end{align}
where
    \bea
    &&F_\pm(\zeta):=\left[j^{'}_{\ell}(\zeta)+\frac{u_\pm j_{\ell}(\zeta)}{\zeta}\right]^2+
    \left(1+ \frac{v_\pm}{\zeta^2}\right)j_{\ell}(\zeta)^{2},
    \label{F=111}\\
    &&u_+:= 1,~~~~~u_-:=\frac{1}{2},~~~~~v_+:=\sT_{1}(\theta)^2,~~~~v_-:= -\left(\ell + \frac{1}{2}\right)^2,
    \label{u-v=pm}
    \eea
$\zeta := k r\eta$, $\mathcal{O}(\kappa^d)$ stands for terms of order $d$ and higher in powers of $\kappa$, and we have used the relation
    \begin{equation}
    \frac{\fn j_{\ell}^{\prime} (\fn x)}{j_\ell (\fn x)} =
	\frac{\eta j_\ell^{\prime} (\zeta)}{j_\ell (\zeta)} - i\zeta
	\left[\frac{j_\ell^{\prime 2} (\zeta)}{j_\ell^{2} (\zeta)}+
	\frac{j_\ell^{\prime} (\zeta)}{\zeta j_\ell (\zeta)} +
	1 - \frac{\ell(\ell + 1)}{\zeta^{2}}\right]\kappa + \mathcal{O}(\kappa^{2}),
	\label{zz-142}
    \end{equation}
which follows from the spherical Bessel equation,
    $j_\ell^{\prime \prime}(\zeta) = - 2 j_\ell^{\prime} (\zeta) /\zeta-
	\left[1- \ell(\ell+1)/\zeta^{2}\right]j_\ell (\zeta)$.

The $\theta$-dependence of $\br u\kt$ complicates the study of its $r$-dependence. This complication can be avoided by realizing that we can fulfil the condition that $\br u\kt$ be concentrated on or near the boundary of the spherical sample by demanding that its average over the solid angle subtended by the sphere, i.e.,
     \be
     \overline{\br u\kt}:=
	\frac{1}{4\pi}\int_{S^2}\br u\kt\,d\Omega=\frac{1}{2}\int_0^\pi\br u\kt\; \sin\theta\, d\theta,
     \label{average-u}
     \ee
has a peak near the boundary of the sphere. In (\ref{average-u}), $S^2$ stands for the unit sphere, and $d\Omega:=\sin\theta\,d\theta\,d\varphi$ is the solid angle element. As shown in the appendix (Eq.~(\ref{uEav=}) below), we can express $\br u\kt$ in terms of an orthogonal set of vector spherical harmonics and use their properties to evaluate the integral over the solid angle in (\ref{average-u}). For $r\leq a$, this gives
	\bea
    \overline{\br u\kt}&=&\frac{\varepsilon_{0}\left|a_0 j_{\ell} (k\fn r)\right|^2}{16\pi }
    \left\{\RE(\fn^{\!\!2}) +
	\frac{1}{k^2r^2}\left[
    \left|\frac{k\fn r\ {j}'_\ell(k\fn r)}{j_\ell(k\fn r)}+1\right|^2+\ell (\ell+1)\right]  \right\}
    \label{uEav=01}\\
    &=&\frac{\varepsilon_{0}\left|a_0 \right|^2 \eta^2 \bar F_{+}(\zeta)}{16\pi}  + \mathcal{O}(\kappa^2),
	\label{uEav=02}
    \eea
where
	\be
    \bar F_{+} (\zeta) := \left[j'_{\ell}(\zeta) +\frac{j_{\ell}(\zeta)}{\zeta}\right]^2 +
	\left[1 + \frac{\ell(\ell+1)}{\zeta^2}\right]j_{\ell}(\zeta)^2.
    \label{F-bar}
    \ee

Because (\ref{uEav=01}) and (\ref{uEav=02}) have the same structure as the corresponding expressions for the cylindrical sample studied in \cite{pra-2013b}, we can make use of the arguments presented in \cite{pra-2013b} to conclude that $\overline{\br u\kt}$ has a peak at the boundary of the sphere, if $\ell\gg 1$ and $\zeta$ is a zero of $j_\ell'$. This leads to a set of surface waves that we label as WGM${'}$. The more conventional whispering gallery modes, that we denote by WGM, correspond to the cases that $\zeta$ is a zero of $j_\ell$, \cite{WGM}.

Notice that the time-averaged energy density $\br u\kt$ of the surface waves that we have constructed is a function of $\ell,m,r$, and $\theta$. The above analysis shows that if we choose $\ell\gg 1$ and $\zeta$ to be the first zero of $j_\ell$ (or $j'_\ell$),  $\br u\kt$ is concentrated in a thin spherical shell of outer radius $a$.

It turns our that, similarly to the cylindrical model studied in \cite{pra-2013b}, neither WGM nor WGM${'}$ of a spherical gain medium is capable of supporting a spectral singularity. In the following we explore the singular gallery modes (SGMs) that by definition support spectral singularities. These are also labeled by an angular mode number, namely $\ell$, and a radial mode number $q$. The latter counts the number of peaks of $\overline{\br u\kt}$ for $\zeta\leq ak\eta$. The peaks (maximum points) of $\overline{\br u\kt}$ as a function of $\zeta$ lie in the intervals bounded by the adjacent positive zeros of $j_\ell$ and $j_\ell'$.

\begin{figure}
    \begin{center}
    \textcolor[rgb]{1.00,1.00,1.00}{A}~\includegraphics[scale=.6]{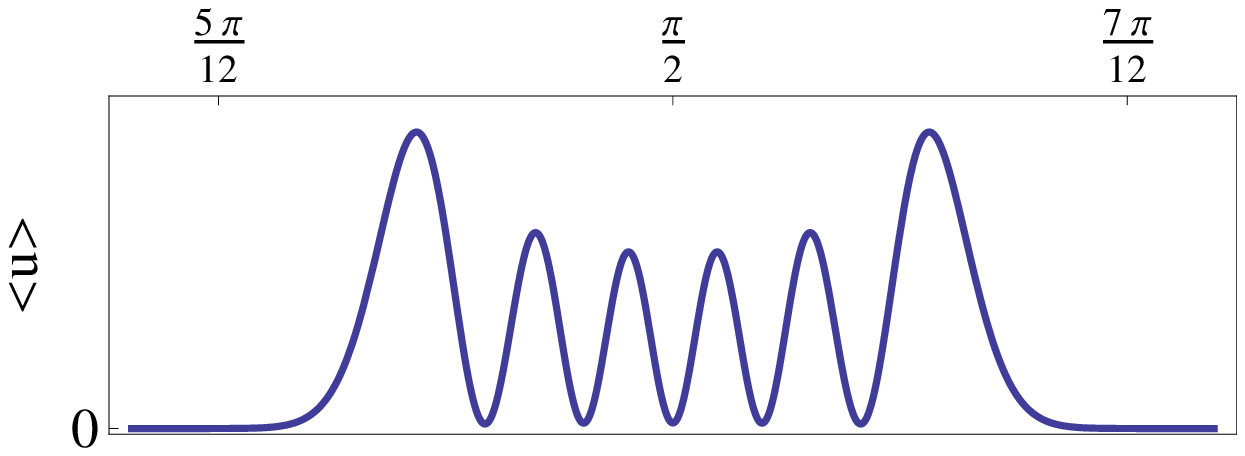}\\
	\includegraphics[scale=.62]{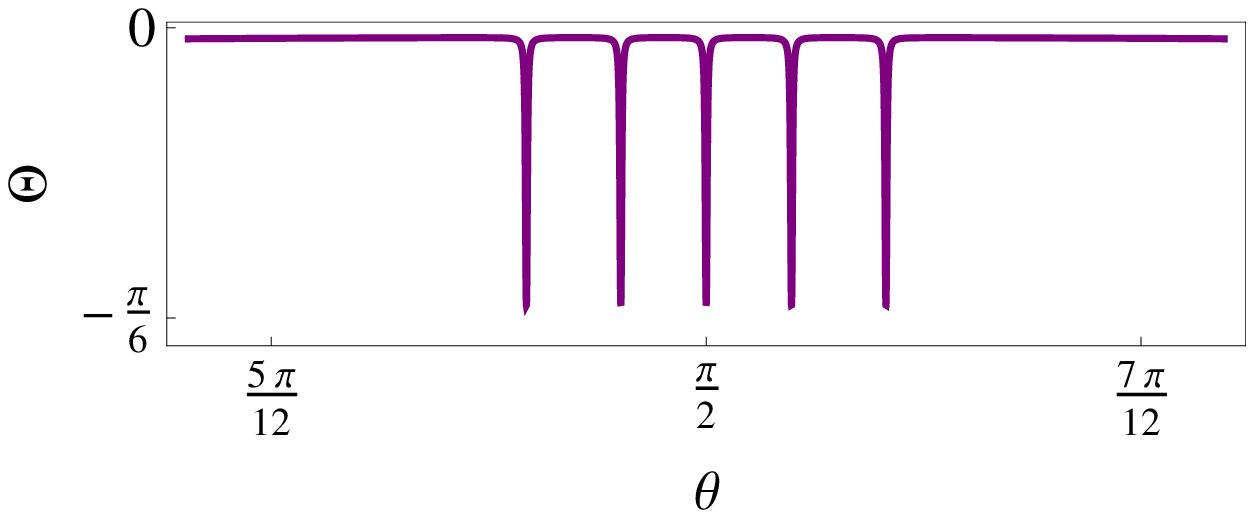}\\
    \caption{(Color online) Graphs of $\br u\kt$ (top) and $\Theta$ (bottom) as a function of $\theta$ for $\ell=350$, $m=345$, $r=25\,\mu{\rm m}$, $\eta=1.82$, $\kappa=10^{-7}$, $\lambda=808~{\rm nm}$, and appropriately chosen $|a_0|$. The minima of $\br u\kt$ coincide with angles $\theta$ at which $\Theta$ differs from zero substantially.}
    \label{fig1}
    \end{center}
    \end{figure}

Before treating the SGMs in more detail, we wish to elaborate on the behavior of the angle between the Poynting vector and the tangent plane to the surface of the sphere, namely, $\Theta$. According to (\ref{Theta=01}) and (\ref{Theta=02}), depending on the value of $\ell$, $m$, and $\theta$, $\sT_{2}(\theta)$ can vanish or take extremely small values, and $\Theta\approx-\pi/2$. Therefore, unlike for a cylindrical sample \cite{pra-2013b}, it is possible that the Poynting vector be normal to the surface of the sphere and at the same time the time-averaged energy density be concentrated on or near its boundary. Figure~\ref{fig1} shows the graphs of $\br u\kt$ and $\Theta$ as a function of $\theta$ for fixed values of $\ell$, $m$, $r$, $\eta=1.81$, and $\kappa=10^{-7}$. As demonstrated by this figure, $\br u\kt$ has $\ell-m$ minima, where it vanishes approximately, and more interestingly these minima coincide with those of $\Theta$. Because $|\Theta|$ takes extremely small values except when $\theta$ is in a close vicinity of the minima of $\Theta$, the condition that $\br u\kt$ attains its maximum near the surface of the spherical sample is compatible with the condition that the wave travels essentially in a direction tangent to this surface.

The TE SGMs are determined using Eq.~(\ref{SSE-eq}) which we can also express as
    \be
    \frac{h_{\ell}^{\prime(1)} (x)}{h_\ell^{(1)} (x)} =
	\frac{\fn j_{\ell}^{\prime} (\fn x)}{j_\ell (\fn x)}.
	\label{TE-spectr}
    \ee
Clearly, this equation does not restrict the range of values of $m$. Therefore, one can consider superpositions of SGMs with different $m$ for the same $\ell$.

Because $\ell^{-1}\ll 1$ and $|\kappa|\ll 1<\eta$, we can gain a better understanding of the consequences of  (\ref{TE-spectr}) by expanding it in powers of $\kappa$ and $\ell^{-1}$. As we show below, this gives a reliable approximation for (\ref{TE-spectr}) that involves trigonometric and hyperbolic functions.

We begin our perturbative treatment of (\ref{TE-spectr}) by using (\ref{zz-142}) to express the right-hand side of (\ref{TE-spectr}) in powers of $\kappa$. Next, we note that the spherical Bessel and Hankel functions are related to the usual Bessel and Hankel functions according to
    \begin{equation}
    \frac{h_{\ell}^{\prime(1)} (x)}{h_\ell^{(1)} (x)} =
	\frac{H_{\nu}^{\prime(1)} (x)}{H_\nu^{(1)} (x)} -\frac{1}{2x},
	~~~~~~~~~~~~~~~~
	\frac{j_\ell^{\prime} (\zeta)}{j_\ell (\zeta)} =
	\frac{J_\nu^{\prime} (\zeta)}{J_\nu (\zeta)}-\frac{1}{2\zeta},
	\label{shift-cyl-to-sph}
    \end{equation}
where $\nu:= \ell + \frac{1}{2}$. Furthermore, we recall that for the typical laser material that we are interested, $\eta>1$ and $|\zeta-\nu|\ll \nu$ so that $x<\nu$. This in turn shows that we can employ Debye's asymptotic expansions \cite{abramowitz},
	\begin{align}
    &J_\nu(\zeta)=\sqrt{\frac{2}{\pi\nu\tan\alpha}}\left[\cos\phi+\cO(\nu^{-1})\right],
	&& J'_\nu(\zeta)=\sqrt{\frac{2}{\pi\nu\tan\alpha}}\left[\sin\phi+\cO(\nu^{-1})\right],
	\label{B2-3}\\
	&H_\nu^{(1)}(x)=\frac{e^\psi-2i e^{-\psi}+\cO(\nu^{-1})}{\sqrt{2\pi\nu\tanh\beta}}, &&
	H_\nu^{(1)\prime}(x)=\sqrt{\frac{\sinh(2\beta)}{4\pi\nu}}
	\left[e^\psi+2i e^{-\psi}+\cO(\nu^{-1})\right],
	\label{B2-4}
	\end{align}
where $\alpha:=\cos^{-1}(\nu/\zeta)\in(0,\frac{\pi}{2})$, $\phi:=\nu(\tan\alpha-\alpha)-\frac{\pi}{4}$, $\beta:=\cosh^{-1}(\nu/x)\in\R^+$, and $\psi:=\nu(\tanh\beta-\beta)$. Finally, we use (\ref{zz-142}) -- (\ref{B2-4}) to compute the real and imaginary parts of (\ref{TE-spectr}). This gives
    \begin{align}
	&\left(\frac{1-4e^{-4\psi}}{1 + 4e^{-4\psi}}\right) \sinh\beta - \eta\tan\phi \approx 0,
	\label{TE-spec-sing1}\\
	&\left(\frac{4e^{-2\psi}}{1 + 4e^{-4\psi}}\right) \sinh\beta+\zeta\left(\sec^{2}\phi - \cos^2\alpha\right)\kappa \approx 0.
	\label{TE-spec-sing2}
	\end{align}
Here $\approx$ stands for the approximate equalities that ignore $\cO(\kappa^2)$ and $\cO(\nu^{-1})$ in (\ref{B2-3}) -- (\ref{B2-4}).

Tables~\ref{table03} gives the numerical values of the physical parameters for the SGMs of a spherical gain medium of radius $a=50 \mu m$, $\ell=600$, and $\eta = 1.8217$. Here $\lambda_{\ell,q}$, $\kappa_{\ell,q}$, and $g_{\ell,q}$ respectively stand for the wavelength, the imaginary part of the complex refractive index, and the gain coefficient required for exciting the SGM labeled by $\ell$ and $q$. The latter satisfies $g_{\ell,q}=-4\pi\kappa_{\ell,q}/\lambda_{\ell,q}$.
    \begin{table}
    \begin{center}
    \begin{tabular}{|c|c|c|c|c|c|}
    \hline
    ${q}$ & $\zeta$ & $\lambda_{\ell,{q}}$ (nm) & $\kappa_{\ell,{q}}$ & $g_{\ell,{q}}\:({\rm cm}^{-1})$ \\
    \hline
    1 & 602.280 & 950.229 & $-1.310 \times 10^{-195}$ & $1.733 \times 10^{-190}$ \\
    2 & 619.951 & 923.144 & $-3.952 \times 10^{-183}$ & $5.379 \times 10^{-178}$ \\
    3 & 631.229 & 906.650 & $-1.957 \times 10^{-175}$ & $2.712 \times 10^{-170}$ \\
    $\vdots$ &$\vdots$&$\vdots$&$\vdots$&$\vdots$\\
    100 & 1083.950 & 527.980& $-6.452 \times 10^{-5}$ & $15.357$ \\
    101 & 1087.746 & 526.137& $-7.931 \times 10^{-5}$ & $18.942$ \\
    102 & 1091.540 & 524.309& $-6.461 \times 10^{-5}$ & $15.486$ \\
    \hline
    \end{tabular}
    \vspace{6pt}
    \caption{The values of $\zeta$, $\lambda_{\ell,{q}}$, $\kappa_{\ell,{q}}$, and $g_{\ell,{q}}$ for spectral singularities of the TE SGMs with $\ell=600$. Here $a=50\:\mu{\rm m}$, $\eta=1.8217$, and $q\leq q_{\rm max}=102$.}
    \label{table03}
    \end{center}
    \end{table}
Table~\ref{table04} gives, for the same sample, the total number of SGMs $q_{\rm max}$ (ignoring the degeneracy related to the choice of the mode label $m$), their spectral range, and the minimum gain coefficient $g_{\rm min}$ required for exciting the SGM corresponding to $q=1$ for several values of $\ell$.
    \begin{table}
    \begin{center}
    \begin{tabular}{|c|c|c|c|c|}
    \hline
    $\ell$ & $q_{\rm max}$ & $\lambda_{\rm min}$ (nm) & $\lambda_{\rm max}$ (nm) &
    $g_{\rm min}\,({\rm cm}^{-1})$ \\
    \hline
    400 & 68 & 787.289 & 1423.418 & $4.007\times 10^{-126}$\\
    500 & 85 & 629.434 & 1139.642 & $2.667\times 10^{-158}$\\
    600 & 102 & 524.309 & 950.229 & $1.733 \times 10^{-190}$\\
    700 & 119 & 449.274 & 814.817 & $1.106 \times 10^{-222}$\\
    \hline
    \end{tabular}
    \vspace{6pt}
    \caption{The values of $q_{\rm max}$ for TE SGMs, the minimum and maximum values of their wavelength, $\lambda_{\rm min}$ and $\lambda_{\rm max}$, and the minimum gain coefficient, $g_{\rm min}$, for $a=50\,\mu{\rm m}$, $\eta=1.8217$, and various values of $\ell$.}
    \label{table04}
    \end{center}
    \end{table}

According to the data depicted in Tables~\ref{table03} and \ref{table04}, the minimum gain required for generating an OSS in the SGMs is a decreasing function of $\ell$ and an increasing function of $q$. In particular, by properly lowering any amount of gain we can excite a SGM of sufficiently large $\ell$. Therefore, similarly to the cylindrical model studied in Ref.~\cite{pra-2013b}, the system lacks a lasing threshold effectively. This shows that we can generate spectral singularities in the surface waves of a micrometer-size spherical gain medium and avoid the technical problems related to maintaining a relatively large and highly uniform gain coefficient throughout a millimeter-size sphere \cite{pla-2011,prsa-2012}.

\subsection{TM Waves}

Pursuing the approach of the preceding subsection and using the formulas given in Table~\ref{table01}, we obtain the following expressions for the time-averaged energy density and Poynting vector for the TM waves.
    \begin{align}
    &\br u\kt=\frac{\mu_0|\sH(r)|^2 \sT_{0}(\theta)^2}{4}\left\{1 + \frac{\RE[\fz(r)^{-1}]}{k^2 r^2} \left[\left|\frac{r\sH'(r)}{\sH(r)}+1\right|^2 + \sT_{1}(\theta)^2 \right]
    \right\},
    \label{TM-u=}\\
    &\br\vec S\kt=-\frac{Z_0|\sH(r)|^2 \sT_{0}(\theta)^2}{2kr\left|\fz(r)\right|^2}\,
    \IM\left\{\fz(r)^* \left[\left(\frac{r\sH'(r)}{\sH(r)}+1\right)\hat r - \sT_{3}(\theta)\,\hat\theta +  i\sT_{2}(\theta)\,\hat\varphi\right] \right\},
    \label{TM-S=}
    \end{align}
where $\sT_{0,1,2}$ are defined by (\ref{T0=}) -- (\ref{T2=}), and
    \[\sT_{3}(\theta):=\frac{T^{(2)}_{\ell m}P^m_{\ell}(\cos\theta)}{\sT_{0}(\theta)^2}=\frac{ \sin(\theta)^2 P^m_\ell(\cos\theta) \partial_\theta P^m_\ell(\cos\theta) }{\sin^2(\theta)\left[\partial_\theta P^m_\ell(\cos\theta)\right]^2+m^2P^m_\ell(\cos\theta)^2
    }.\]

Again inside the sphere, where $\fz(r)=\fn^2$, $\sH(r)=a_0 j_\ell(k\fn r)$, and Eqs.~(\ref{TM-u=}) and (\ref{TM-S=}) take the form
    \begin{align}
    &\br u\kt=\frac{\mu_0\big|a_0 j_\ell(k\fn r)\big|^2 \sT_{0}(\theta)^2}{4}\left\{1 + \frac{\RE(\fn^{\!\!-2})}{k^2 r^2}\left[\left|\frac{k\fn r j'_\ell(k\fn r)}{j_\ell(k\fn r)} + 1\right|^2 + \sT_{1}(\theta)^2 \right]
    \right\},
    \label{TM-u=3}\\
    &\br\vec S\kt= -\frac{Z_0\big|a_0 j_\ell(k\fn r)\big|^2 \sT_{0}(\theta)^2}{2kr\left|\fn^{2}\right|^2}\,
    \IM\left\{ \fn^{\!\!2*} \left[\left(\frac{k\fn r j'_\ell(k\fn r)}{j_\ell(k\fn r)}+1\right)\hat r -  \sT_{3}(\theta)\,\hat\theta + i\sT_{2}(\theta)\,\hat\varphi\right] \right\}.
    \label{TM-S=3}
    \end{align}
Expressing the right-hand side of these relations in terms of $\eta$ and $\kappa$ and realizing that
$\kappa \ll 1<\eta$, we can show that
     \begin{align}
     \br u\kt &= \frac{\mu_{0}\left|a_0\right|^2 \sT_{0}(\theta)^2 F_+(\zeta)}{4} +\mathcal{O}(\kappa^2),\\
      \Theta &= \tan^{-1}\left[\frac{\kappa}{\eta \sT_{2}(\theta)}\frac{\zeta^2 F_{-}(\zeta)}{j^2_{\ell}(\zeta)}\right]+\mathcal{O}(\kappa^2),
     \end{align}
where $F_\pm$ are given by (\ref{F=111}) and (\ref{u-v=pm}) except that now $u_-:= 3/2$.

OSSs in the TM modes are determined by Eq.~(\ref{SSM-eq}) that we express as
    \be
    \frac{h_{\ell}^{\prime(1)} (x)}{h_\ell^{(1)} (x)} +
    \frac{1}{x}\left(1-\frac{1}{\fn^2}\right)=
    \frac{j_{\ell}^{\prime} (\fn x)}{\fn j_\ell (\fn x)}.
    \label{TM-spectr}
    \ee
In view of Eqs.~(\ref{shift-cyl-to-sph}), the identity
    \begin{equation}
    \frac{j_{\ell}^{\prime} (\fn x)}{\fn j_\ell (\fn x)} =
    \frac{j_\ell^{\prime} (\zeta)}{\eta j_\ell (\zeta)} - \frac{i\zeta}{\eta^{2}}\left[\frac{j_\ell^{\prime 2} (\zeta)}{j_\ell^{2} (\zeta)}+\frac{3 j_\ell^{\prime} (\zeta)}{\zeta j_\ell (\zeta)} + 1 - \frac{\ell(\ell + 1)}{\zeta^{2}}\right]\kappa + \mathcal{O}(\kappa^{2}),
    \end{equation}
and Debye's asymptotic expansions (\ref{B2-3}) and (\ref{B2-4}), we find the following expressions for the real and imaginary parts of (\ref{TM-spectr}) and the angle $\Theta$.
    \begin{align}
	&\left(\frac{1-4e^{-4\psi}}{1 + 4e^{-4\psi}}\right) \sinh\beta - \frac{\tan\phi}{\eta}
    +\frac{\eta^{2}-1}{2\eta\zeta}\approx 0,
	\label{TM-spec-sing1}\\
	&\left(\frac{4e^{-2\psi}}{1 + 4e^{-4\psi}}\right) \sinh\beta+\frac{\zeta}{\eta^{2}}
	\left[\left(\tan\phi +\frac{1}{\zeta}\right)^{2} +
    1-\frac{(\nu^{2}+2)}{\zeta^{2}}\right]\kappa + \frac{2\kappa}{\zeta\eta^2} \approx 0,
	\label{TM-spec-sing2}\\
	&\Theta \approx \tan^{-1}\left\{\frac{\kappa}{\eta \sT_{2}(\theta)}
    \left[(\zeta \tan \phi + 1)^2 +
    \zeta^2 - (\ell + \frac{1}{2})^2\right]\right\}.
    \label{TM-angle}
     \end{align}

We have used Eqs.~(\ref{TM-spec-sing1}) and (\ref{TM-spec-sing2})  to determine the physical parameters associated with the TM SGMs of our spherical model. We summarize the results in Tables~\ref{table5} and \ref{table6} which are the TM-analogs of Tables~\ref{table03} and \ref{table04}. They lead to a similar conclusion regarding the lack of a lasing threshold for the spherical sample under consideration.
   \begin{table}
    \begin{center}
    \begin{tabular}{|c|c|c|c|c|c|}
    \hline
    ${q}$ & $\zeta$ & $\lambda_{\ell,{q}}$ (nm) & $\kappa_{\ell,{q}}$ & $g_{\ell,{q}}\:({\rm cm}^{-1})$ \\
    \hline
    1 & 617.723 & 926.473 & $-3.661 \times 10^{-185}$ & $4.965 \times 10^{-180}$ \\
    2 & 629.447 & 909.217 & $-4.188 \times 10^{-177}$ & $5.788 \times 10^{-172}$ \\
    3 & 639.230 & 895.302 & $-1.504 \times 10^{-170}$ & $2.111 \times 10^{-165}$ \\
    $\vdots$ &$\vdots$&$\vdots$&$\vdots$&$\vdots$\\
    99 & 1083.756 & 528.075& $-1.964 \times 10^{-4}$ & $46.726$ \\
    100 & 1087.608 & 526.204& $-2.537 \times 10^{-4}$ & $60.598$ \\
    101 & 1091.469 & 524.343& $-2.149 \times 10^{-4}$ & $51.510$ \\
    \hline
    \end{tabular}
    \vspace{6pt}
    \caption{The values of $\lambda_{\ell,{q}}$, $\kappa_{\ell,{q}}$, and $g_{\ell,{q}}$ for the TM-SGMs with $\ell=600$. Here $a=50\:\mu{\rm m}$, $\eta=1.8217$, and ${q}$ takes values between 1 and 101.}
    \label{table5}
    \end{center}
    \end{table}

    \begin{table}
    \begin{center}
    \begin{tabular}{|c|c|c|c|c|}
    \hline
    $\ell$ & $q_{\rm max}$ & $\lambda_{\rm min}$ (nm) & $\lambda_{\rm max}$ (nm) &
    $g_{\rm min}\,({\rm cm}^{-1})$ \\
    \hline
    400 & 67 & 787.393 & 1377.045 & $4.428\times 10^{-117}$\\
    500 & 84 & 629.490 & 1107.542 & $1.675\times 10^{-148}$\\
    600 & 101 & 524.343 & 926.473 & $4.965 \times 10^{-180}$\\
    700 & 118 & 449.296 & 796.404 & $1.230 \times 10^{-211}$\\
    \hline
    \end{tabular}
    \vspace{6pt}
    \caption{Values of $q_{\rm max}$, $\lambda_{\rm min}$, $\lambda_{\rm max}$, and $g_{\rm min}$ for the TM SGMs with different $\ell$. Again $a=50\,\mu{\rm m}$ and $\eta=1.8217$.}
    \label{table6}
    \end{center}
    \end{table}

\section{Singular Gallery Modes in the Presence of Dispersion}

In Section~3 we explored the TE and TM SGMs for a spherical gain medium whose refractive index $\fn$ was assumed not to depend on the wavelength. In this section we generalize our treatment of SGM by taking into account the dispersion effects.

First, we recall that among the physical parameters $x, \eta, \kappa$ and $\nu$ entering the description of the SGMs of our model, $x$ and $\kappa$ depend on the radius $a$, the wavelength $\lambda$, and the gain coefficient $g$ of the medium according to \cite{silfvast}
    \be
    x=ak=\frac{2\pi a}{\lambda},~~~~~~~~~~~\kappa=-\frac{\lambda g}{4\pi}.
    \label{x-kappa=}
    \ee
Next, suppose that the gain material filling the sphere consists of an optically active medium that is obtained by doping a host medium of refraction index $n_0$ and modeled by a two-level atomic system with lower and upper level population densities $N_l$ and $N_u$, resonance frequency $\omega_0$, damping coefficient $\gamma$, and the dispersion relation:
    \be
    \fn^2= n_0^2-
    \frac{\hat\omega_p^2}{\hat\omega^2-1+i\hat\gamma\,\hat\omega},
    \label{epsilon}
    \ee
where $\hat\omega:=\omega/\omega_0$, $\hat\gamma:=\gamma/\omega_0$,
$\hat\omega_p:=(N_l-N_u)e^2/(m_e\varepsilon_0\omega_0^2)$, $e$ is
electron's charge, and $m_e$ is its mass. Denoting by $\kappa_0$ the
imaginary part of $\fn$ at the resonance wavelength, $\lambda_0:=2\pi c/\omega_0$,
we have $\hat\omega_p^2=2n_0\hat\gamma\kappa_0+\cO(\kappa_0^2)$, \cite{pra-2011a}.
Inserting this relation in (\ref{epsilon}), using
$\fn=\eta+i\kappa$, and neglecting $\cO(\kappa_0^2)$, we obtain \cite{pla-2011}
    \begin{align}
    &\eta\approx n_0+\kappa_0f_1(\hat\omega),
    &&\kappa\approx\kappa_0f_2(\hat\omega),
    \label{eqz301}
    \end{align}
where $f_1(\hat\omega):= \hat\gamma(1-\hat\omega^2)/[(1-\hat\omega^2)^2+
    \hat\gamma^2\hat\omega^2]$ and $f_2(\hat\omega):= \hat\gamma^2\hat\omega/[(1-\hat\omega^2)^2+
    \hat\gamma^2\hat\omega^2]$.
Next, we recall from (\ref{x-kappa=}) that $\kappa_0$ is related to the gain coefficient at resonant wavelength, $g_0$, according to $\kappa_0=-\lambda_{0}g_0/(4\pi)$. Substituting this equation in (\ref{eqz301}) and using the resulting relations together with (\ref{x-kappa=}) in
Eqs.~(\ref{TE-spectr}) and (\ref{TM-spectr}), we can determine the values of $\lambda$ and $g_{0}$ associated with OSSs.

For definiteness we consider a sphere of radius $a = 50\,\mu{\rm m}$ made out of Nd:YAG crystals with the following specifications \cite{silfvast,Ndyag}.
    \begin{align}
    &n_0=1.8217, &&\lambda_0=808\,{\rm nm}, &&
    \hat\gamma=0.003094, &&g_0\leq 0.359\,{\rm cm}^{-1}.
    \label{specific}
    \end{align}
In the following we summarize the results of our numerical investigation of the OSSs in the TE SGMs of the gain medium (\ref{specific}). These make use of the exact equation for the OSSs, i.e., (\ref{TE-spectr}) and (\ref{TM-spectr}), and the dispersion relations (\ref{eqz301}).

\begin{figure}
    \begin{center}
	\includegraphics[scale=.5]{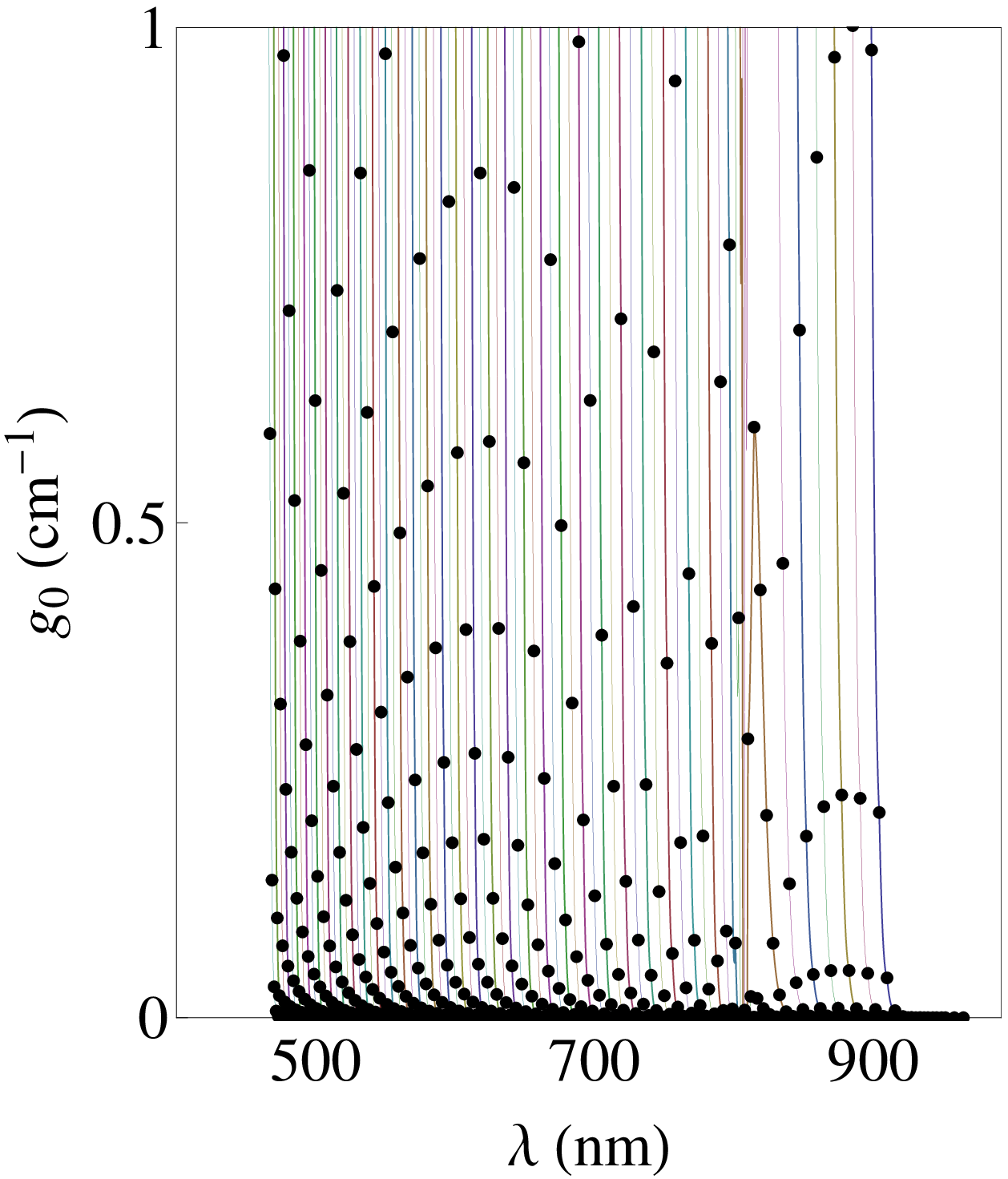}\hspace{1cm}
	\includegraphics[scale=.54]{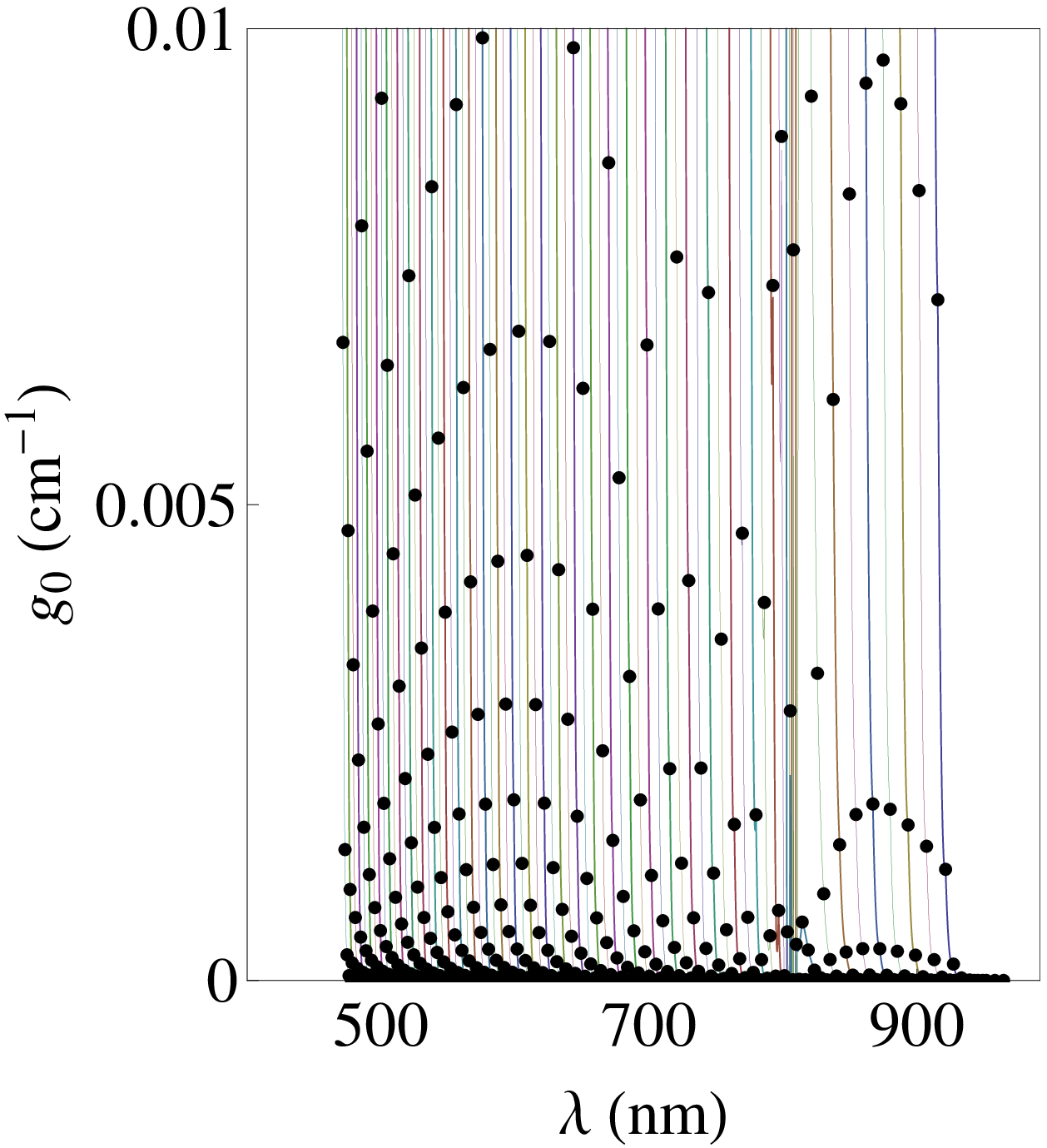}\\
    \caption{(Color online) OSSs associated with the TE SGMs of a sample of radius $50\,\mu{\rm m}$ consisting of the Nd:YAG crystals (\ref{specific}). Each curve corresponds to a particular value of $\ell$ that ranges from 375 (the rightmost curve) to 710 (the leftmost curve) in increments of 5. The displayed dots represent OSSs. Essentially identical curves are obtained for OSSs in the TM SGMs.}
    \label{fig2}
    \end{center}
    \end{figure}

\begin{figure}
    \begin{center}
	\includegraphics[scale=.43]{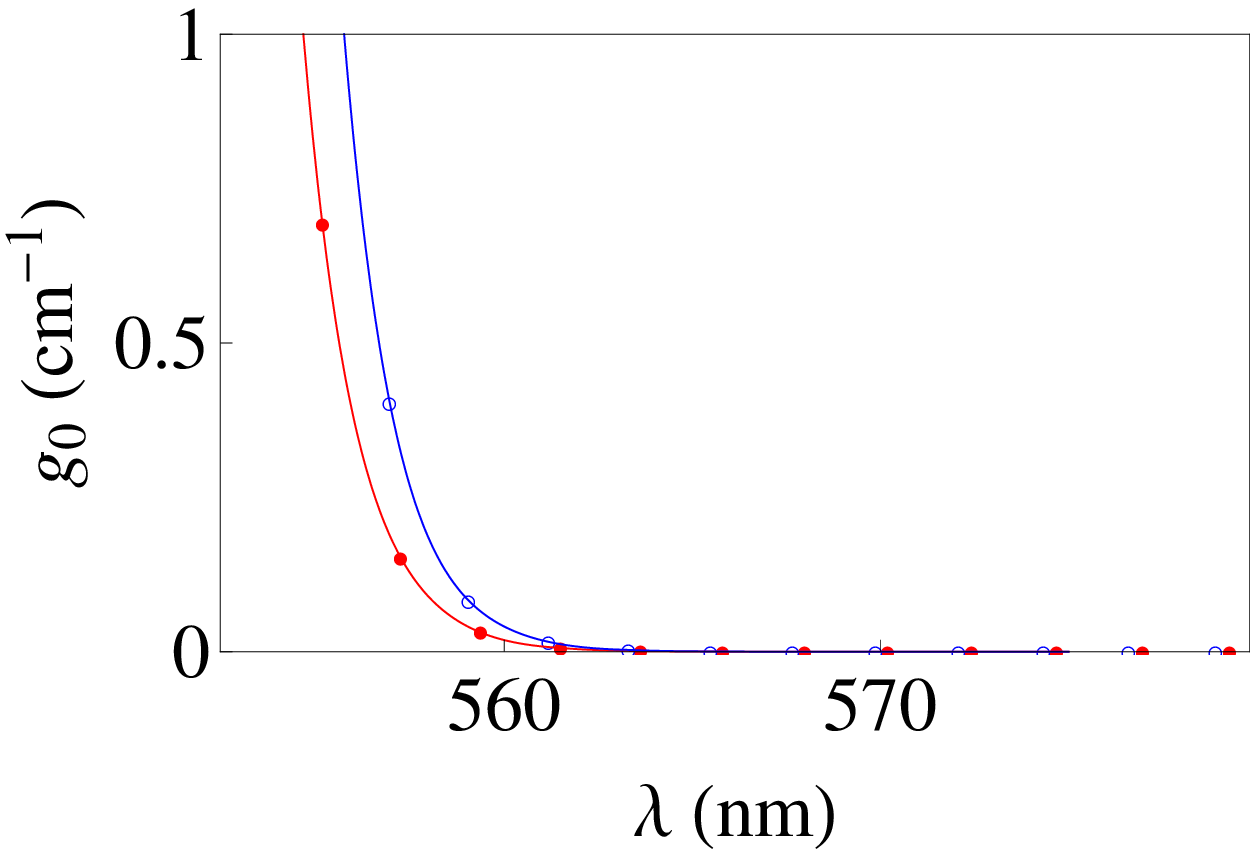}\hspace{.5cm}
	\includegraphics[scale=.46]{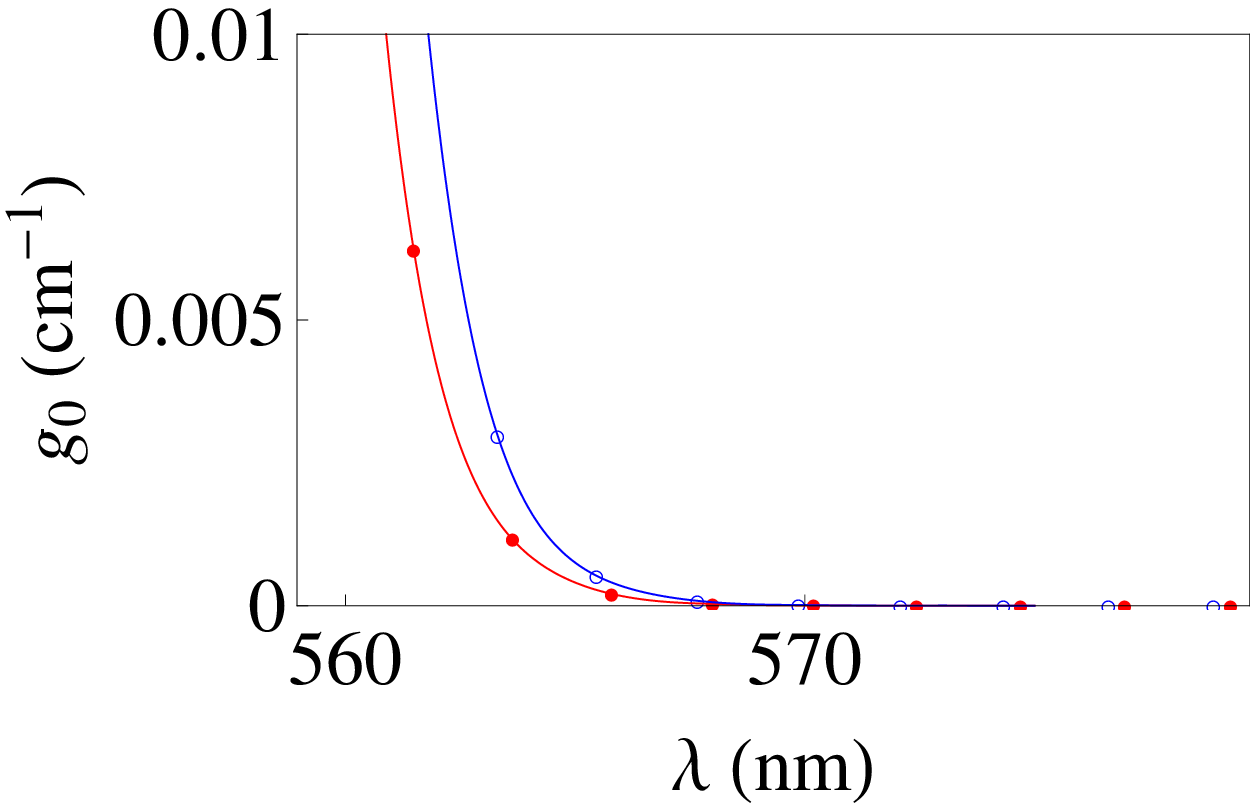}\\
    \caption{(Color online) A comparison of the OSS in the TE (the red curve on the left) and TM (the blue curve on the right) SGMs for the same sample as in Figure~\ref{fig2}. The displayed dots represent the spectral singularities.}
    \label{fig3}
    \end{center}
    \end{figure}

Figures~\ref{fig2} and ~\ref{fig3} show the location of the OSSs in the $\lambda$-$g_0$ plane for TE and TM SGMs with various choices of $\ell$.
The presence of OSSs for extremely small gain coefficients confirms our expectation that surface waves can support OSSs for very small values of the radius.

A graphical study of the $\ell$-dependence of gain coefficient $g_0$ that is required for exciting SGMs shows that $g_0$ is a decreasing function of $\ell$ that tends to zero as $\ell$ becomes arbitrarily large. This agrees with the perturbative results we have listed in Tables~\ref{table04} and~\ref{table6} as well as the results reported in Ref.~\cite{pra-2013b} for the cylindrical gain media.

\begin{figure}
	\begin{center}
    \includegraphics[scale=.50]{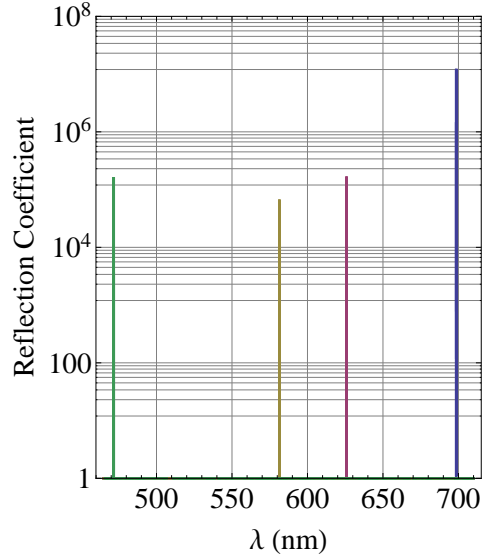}
	\caption{(Color online)  Logarithmic plots of the TE reflection coefficient $|R|^2$ as a function of $\lambda$ for $a=50\:\mu{\rm m}$, $g_0= 6.682\times 10^{-3}\:{\rm cm}^{-1}$, and $\ell= 485,\, 540,\, 580,\, 710$ from right to left, respectively.}
    \label{fig4}
    \end{center}
    \end{figure}

Our numerical studies also confirm the observation that SGMs have a divergent quality factor, and that given their abundance, we can simultaneously excite several SGMs using the same amount of gain. Figure~\ref{fig4} shows the graph of the reflection coefficient $|R|^2$ as a function of $\lambda$ for a situation in which this scenario holds effectively.
The peaks represent four different OSSs that arise for $g_0\approx 6.682\times 10^{-3}\,{\rm cm}^{-1}$. Table~\ref{table7} gives the corresponding $\ell$, $g_0$, and $\lambda$ values.
    \begin{table}
    \begin{center}
    \begin{tabular}{|c|c|c|}
    \hline
    $\ell$ & $g_0\:({\rm cm}^{-1})$ & $\lambda$ (nm) \\
    \hline
    485 & $6.67824\times 10^{-3}$ & 698.540903\\
    540 & $6.71443\times 10^{-3}$ & 625.862146\\
    580 & $6.63064\times 10^{-3}$ & 581.419242\\
    710 & $6.70446\times 10^{-3}$ & 471.648017\\
    \hline
    \end{tabular}
    \vspace{6pt}
    \caption{Values of $\ell$, $g_0$, and $\lambda$ corresponding to the OSSs depicted in Figure~\ref{fig4}.}
    \label{table7}
    \end{center}
    \end{table}

\section{Concluding Remarks}

Singular gallery modes are surface modes that support spectral singularities. They share most of the properties of the whispering gallery modes but unlike them have a divergent quality factor. In this article we have explored the TE and TM singular gallery modes of a homogeneous spherical gain medium. These are determined by a set of rather complicated mathematical expressions that have similar structure to those of an infinite cylindrical gain medium \cite{pra-2013b}. The main distinguishing factor is that for a spherical gain medium the time-averaged energy density $\br u\kt$ depends on both the radial and polar coordinates, $r$ and $\theta$. The same is true for the angle $\Theta$ between the time-averaged Poynting vector and the tangent plane to the surface of the sphere.

For a surface wave, we expect that $\br u\kt$ has peaks on or near the surface of the sphere while $|\Theta|$ is very small. A careful study of $\br u\kt$ and $\Theta$ for a surface wave with mode labels $\ell$ and $m$ shows that, for sufficiently large $\ell$, $\Theta$ deviates from zero noticeably only within extremely narrow intervals around its minimum points that are $\ell-m$ in number. These points that are symmetrically located about $\theta=\pi/2$ happen to coincide with the minima of $\br u\kt$ at which $\br u\kt$ vanishes approximately. This observation shows that the conditions on $\br u\kt$ and $\Theta$ that characterize the surface waves of the sphere and used in the study of the whispering and singular gallery modes are indeed consistent.

The condition that a surface wave supports a spectral singularity and hence has a divergent quality factor does not involve the mode label $m$. Therefore, once one can generate a spectral singularity in a surface mode with mode labels $\ell_\star$ and $m_\star$, there will appear spectral singularities with mode label $\ell_\star$ and arbitrary $m$ in the range $-\ell_\star,-\ell_\star+1,\cdots,\ell_\star$. This shows that the singular gallery modes can be uniquely determined by the angular mode number $\ell$ and a radial mode number $q$ that counts the peaks of the time- and solid-angle-averaged energy density $\overline{\br u\kt}$ as we increase $r$ from $0$ to $a$.

Our perturbative, numerical, and graphical study of singular gallery modes show that for any amount of gain, the system includes singular gallery modes corresponding to smaller gains. Hence, similarly to the cylindrical sample of Ref.~\cite{pra-2013b}, a spherical gain medium lacks a lasing threshold, effectively.

Another common feature of the cylindrical and spherical models is that the abundance of singular gallery modes allows to consider situations that the system possesses distinct spectral singularities for approximately the same amount of gain. This means that one can excite different singular gallery modes simultaneously and make the system emit non-monochromatic waves.

Finally, we wish to stress that our treatment of the surface waves of a spherical medium  differs from the standard approach used in the study of the whispering gallery modes \cite{WGM} in that the latter employs the uniform asymptotic expansion of the spherical Bessel and Hankel functions \cite{abramowitz} and yields large-$\ell$ asymptotic series involving the Airy function and its derivative. We instead use Debye's asymptotic expansion which is less restrictive. In fact, it turns out that the uniform asymptotic expansion is not capable of characterizing all the singular gallery modes.
\vspace{.3cm}

\noindent \textbf{{Acknowledgments:}} We wish to thank Ali Serpeng\"{u}zel and Aref Mostafazadeh for useful discussions. This work has been supported by the Scientific and Technological Research Council of Turkey (T\"UB\.{I}TAK) in the framework of the project no: 110T611, and by the Turkish Academy of Sciences (T\"UBA).

\section*{Appendix}

A well-known tool in the study of the surface waves of a spherical medium is the vector spherical harmonics (VSH) \cite{VSH}. In the following we summarize some of their properties and use them to derive useful expressions for the time-averaged energy density and Poynting vector.

VSHs are defined by
        \bea
        \vec Y_{\ell m}&:=&Y_{\ell m} \hat{r}= b_{\ell m} P_{\ell}^{m} (\cos \theta) e^{im\varphi} \hat{r}\nn\\
        \vec \Psi_{\ell m}&:=& r\vec\nabla Y_{\ell m}=b_{\ell m} e^{im\varphi}\left[T_{\ell m}^{(2)}(\theta)\hat{\theta} + i T_{\ell m}^{(1)}(\theta) \hat{\varphi}\right],\nn\\
        \vec \Phi_{\ell m}&:=& \vec{r} \times \vec\nabla Y_{\ell m}=b_{\ell m} e^{im\varphi}\left[T_{\ell m}^{(2)}(\theta)\hat{\varphi} - i T_{\ell m}^{(1)}(\theta) \hat{\theta}\right],
        \label{append1}
        \eea
where $Y_{\ell m}(\theta, \varphi)$ is the scalar spherical harmonics, $b_{\ell m} = \sqrt{\frac{2\ell + 1}{4\pi}\frac{(\ell - m)!}{(\ell + m)!}}$, and $T_{\ell m}^{(1,2)}$ are given by (\ref{T1-2}). They satisfy the following orthogonality relations \cite{VSH}.
    \begin{align*}
    &\vec Y_{\ell m}\cdot\vec \Psi_{\ell m}=\vec Y_{\ell m}\cdot\vec \Phi_{\ell m}=\vec \Psi_{\ell m}\cdot\vec \Phi_{\ell m}=0,\\
    &\int\vec Y_{\ell m}\cdot \vec Y^*_{\ell'm'}\,\mathrm{d}\Omega= \delta_{\ell\ell'}\delta_{mm'},\\
    &\int\vec \Psi_{\ell m}\cdot \vec \Psi^*_{\ell'm'}\,\mathrm{d}\Omega =
    \int\vec \Phi_{\ell m}\cdot \vec \Phi^*_{\ell'm'}\,\mathrm{d}\Omega=
    \ell(\ell+1)\delta_{\ell\ell'}\delta_{mm'},\\
    &\int\vec Y_{\ell m}\cdot \vec \Psi^*_{\ell'm'}\,\mathrm{d}\Omega =
    \int\vec Y_{\ell m}\cdot \vec \Phi^*_{\ell'm'}\,\mathrm{d}\Omega=
    \int\vec \Psi_{\ell m}\cdot \vec \Phi^*_{\ell'm'}\,\mathrm{d}\Omega= 0.
    \end{align*}

In view of the identity:
    \be
    \vec\nabla \times \left[f(r)\vec \Phi_{\ell m}\right] = -\frac{\ell(\ell+1)}{r} f(r)\vec Y_{\ell m}-\tilde f(r)\vec \Psi_{\ell m},\notag
    \ee
that holds for every differentiable scalar-valued function $f$, and the fact that $\vec X_{\ell m}=-i \vec \Phi_{\ell m}/\sqrt{\ell(\ell+1)}$, we can express the TE-fields and the corresponding time-averaged energy density and Poynting vector in the form
     \bea
     \vec E &=& -i\frac{\sE(r)}{\sqrt{\ell(\ell+1)}}\,\vec \Phi_{\ell m},\nn\\
     \vec H &=& \frac{1}{Z_{0} k\sqrt{\ell(\ell+1)}}\left[\frac{\ell(\ell+1)}{r}\sE(r)\vec Y_{\ell m} + \tilde \sE(r) \vec \Psi_{\ell m}\right],\nn\\
    \br u\kt&=& \frac{\epsilon_{0}\left|\sE(r)\right|^2}{4}
    \left\{\RE(\fz) |\vec \Phi_{\ell m} |^2 +
    \frac{1}{k^2 r^2}\left[\left|\frac{r\sE'(r)}{\sE(r)}+1\right|^2
     |\vec \Psi_{\ell m} |^2 +
    \ell^2 (\ell+1)^2 |\vec Y_{\ell m} |^2 \right]\right\},\nn\\
    \br\vec S\kt &=& \frac{\left|\sE(r)\right|^2}{2Z_{0}kr} \IM\left[\frac{r\sE'(r)}{\sE(r)}\frac{\left|\vec \Phi_{\ell m}\right|^2}{\ell(\ell+1)}\hat{r}
    -Y^*_{\ell m}\vec\Psi_{\ell m}\right].\nn
    \eea
We can use the above orthogonality properties of VSHs to compute the average of $\br u\kt$ over the
solid angle subtended by the sphere that we introduced in (\ref{average-u}). The result is
    \be
    \overline{\br u\kt}= \frac{\epsilon_{0}\left|\sE(r)\right|^2}{16 \pi}
    \left\{\RE(\fz) + \frac{1}{k^2 r^2}\left[\left|\frac{r\sE'(r)}{\sE(r)} + 1\right|^2 + \ell(\ell+1)\right] \right\}.
    \label{uEav=}
    \ee
Similarly we find for the TM-fields:
    \bea
    \vec E &=& -\frac{Z_{0}}{k\fz(r)\sqrt{\ell(\ell+1)}}
    \left[\frac{\ell(\ell+1)}{r}\sH(r)\vec Y_{\ell m} + \tilde \sH(r)\vec\Psi_{\ell m}\right],\nn\\
    \vec H &=& -i\frac{\sH(r)}{\sqrt{\ell(\ell+1)}}\vec \Phi_{\ell m},\nn\\
    \br u\kt &=& \frac{\mu_{0}\left|\sH(r)\right|^2}{4}
    \left\{ |\vec \Phi_{\ell m} |^2 + \frac{\RE[\fz(r)]}{k^2 r^2 \left|\fz(r)\right|^2} \left[\left|\frac{r\sH'(r)}{\sH(r)}+1\right|^2  |\vec \Psi_{\ell m} |^2 + \ell^2 (\ell+1)^2 |\vec Y_{\ell m} |^2 \right] \right\},\nn\\
    \br\vec S\kt &=& \frac{Z_{0}\left|\sH(r)\right|^2}{2kr\left|\fz(r)\right|^2}
	\IM \left\{\fz(r)^* \left[Y_{\ell m}\vec \Psi_{\ell m}^*
	-\frac{r\sH'(r)}{\sH(r)}
	\frac{ |\vec \Psi_{\ell m} |^2}{\ell(\ell+1)}\hat{r}\right]\right\},\nn\\
    \overline{\br u\kt}&=&\frac{\mu_{0}\left|\sH(r)\right|^2}{16\pi}
    \left\{1 + \frac{\RE[\fz(r)]}{k^2 r^2 \left|\fz(r)\right|^2} \left[\left|\frac{\sH'(r)}{\sH(r)}+1\right|^2 +\ell (\ell+1) \right] \right\}.
    \label{uav=}
    \eea

Using $\fz(r) = \fn^2$, $\sE(r) = \sH(r)= a_0 j_{\ell} (k\fn r)$ for $r < a$, and the above expressions for the VSHs, we have checked that the above relations are in agreement with those listed in the text. In particular, for $r\leq a$,
    \be
    \overline{\br u\kt}=\left\{\begin{array}{ccc}
    \displaystyle\frac{\epsilon_{0}\left|a_0 \right|^2 \eta^2 \bar F_{+}(\zeta)}{16 \pi}  + \mathcal{O}(\kappa^2) & {\rm for} & \rm TE~mode,\\[12pt]
    \displaystyle \frac{\mu_{0}\left|a_0 \right|^2 \bar F_{+}(\zeta)}{16 \pi}  + \mathcal{O}(\kappa^2) & {\rm for} & \rm TM~mode,
    \end{array}\right.
    \label{equation09}
    \ee
where $\bar F_+$ is given by (\ref{F-bar}) and $\zeta := k \eta r$.

\end{document}